\documentclass[10pt,nofootinbib]{article}
\pdfoutput=1
\usepackage[english]{babel}
\usepackage{cite,amsmath,amssymb,amsbsy,amstext, amsthm, simplewick}
\usepackage{hyperref}
\usepackage{graphicx}
\usepackage{amsfonts}
\usepackage{amssymb} 
\usepackage[small]{caption}
\usepackage{upgreek}
\usepackage{exscale,relsize}
\usepackage[makeroom]{cancel}
\usepackage{soul}
\usepackage{lscape}
\usepackage{indentfirst}
\usepackage{latexsym}
\usepackage{multirow}
\usepackage{tabls}
\usepackage{epsfig}
\usepackage{bm}
\usepackage{mathrsfs}
\usepackage{comment}
\usepackage{xspace}
\usepackage[usenames,dvipsnames]{color}
\allowdisplaybreaks[1]

\allowdisplaybreaks[1]

\RequirePackage{color}


\numberwithin{equation}{section}
\def\nn{{\nonumber}}

\def\beq{\begin{equation}}
\def\eeq{\end{equation}}
\def\bea{\begin{eqnarray}}
\def\eea{\end{eqnarray}}

\def\0{{\boldsymbol 0}}

\def\bk{{\boldsymbol{k}}}
\def\bq{{\boldsymbol{q}}}
\def\bp{{\boldsymbol{p}}}

\def\bx{{\boldsymbol{x}}}

\def\bea{\begin{eqnarray}}
\def\eea{\end{eqnarray}}
\def\nn{{\nonumber}}

\def\bk{{\boldsymbol{k}}}
\def\bp{{\boldsymbol{p}}}
\def\bq{{\boldsymbol{q}}}

\def\bx{{\boldsymbol{x}}}

\def\br{{\boldsymbol{r}}}

\def\bd{{\boldsymbol{d}}}

\def\bA{{\boldsymbol{A}}}

\def\addCMU{ \small Department of Physics and Astronomy\\ Carnegie Mellon University, Pittsburgh, Pennsylvania 15213, USA}
\def\addICTP{\small  ICTP South American Institute for Fundamental Research\\ Rua Dr. Bento Teobaldo Ferraz 271, 01140-070 S\~ao Paulo, SP Brazil}
\begin{document}

\begin{titlepage}

\setcounter{page}{1} \baselineskip=15.5pt \thispagestyle{empty}

\begin{center}

{\fontsize{20}{28}\selectfont  \sffamily On the Apparent Ambiguities in the\\ [0.2cm]  Post-Newtonian Expansion for Binary Systems}
\end{center}

\begin{center}
\fontsize{12}{16} \selectfont \sffamily Rafael A.~Porto$^{1}$ and Ira Z. Rothstein$^{2}$\\
\end{center}
\begin{center}
\textsl{$^1$ \addICTP}\vskip 4pt

\textsl{$^2$ \addCMU}\end{center}

\vspace{1.2cm}
\hrule \vspace{0.3cm}
We discuss the source of the apparent ambiguities arising in the calculation of the
dynamics of binary black holes within the Post-Newtonian framework. Divergences appear in both the near and far zone calculations, and may be of either ultraviolet (UV) or infrared (IR) nature. The effective field theory (EFT) formalism elucidates the origin of the singularities which may introduce apparent ambiguities. In~particular, the only (physical) `ambiguity parameters' that necessitate a matching calculation correspond to unknown finite size effects, which first appear at fifth Post-Newtonian (5PN) order for non-spinning bodies. We demonstrate that the ambiguities linked to IR divergences in the near zone, that plague the recent derivations of the binding energy at 4PN order, both in the Arnowitt, Deser, and Misner
(ADM) and `Fokker-action' approach, can be resolved by implementing the so-called {\it zero-bin} subtraction in the EFT framework.  The procedure yields ambiguity-free results without the need of additional information beyond the PN expansion.\vskip 10pt
\hrule
\end{titlepage}

\section{Introduction}

The calculation of high accuracy Post-Newtonian (PN) templates for binary inspirals (for extensive reviews see \cite{blanchet,Buoreview,Tiec}) plays a key role in the data analysis program in this nascent era of gravitational wave (GW) science \cite{Abadie:2010cf,Abbott:2016blz,2016htt,2016pea,Sesana}. The PN formalism is a systematic expansion in the relative velocity, $v$, of the constituents during the early stages of the inspiral, for which $v/c \ll 1$.  The expansion parameter also serves to separate out the relevant length scales in the problem: The orbital radius $r$ and the wavelength of the radiation $\lambda_{\rm rad} \sim r/v$. In addition, there is a third scale, namely the  typical radius of the constituents,~$R$, which is, in general, independent of $r$ and $v$. The starting point of a perturbative approach corresponds to a multipole expansion where extended objects, such as neutron stars or black holes, are treated as localized sources. In principle, this is logically independent of the PN expansion, and it is valid even in the relativistic limit, as long as the typical wavelength (or frequency) of the perturbation, $\lambda$, obeys $\lambda \gg R$. For a bound state of compact objects ($R \sim 2G_Nm),\footnote{For objects other than black holes, the compactness: $c \equiv R/(2G_Nm)$ depends on the equation of state. Therefore, while finite size effects will be organized in powers of $v$, there could be large coefficients, e.g. $c^5$ for neutron stars, that may enhance the naive power counting.}$ the virial theorem relates the size of the bodies to the orbital radius, $R/r \sim v^2$, and therefore all relevant scales in the problem are organized in powers of $v$.\vskip 4pt 

In the PN formalism, after the constituents are replaced by point-like sources, the calculations are then separated into {\it regions}, the near (or potential) zone entailing quasi-instantaneous modes $(p^0 \ll |\bp|$) of the gravitational field varying on the scale of the bound state radius, and the far (or radiation) zone, where modes propagate ($p^0 \sim |\bp|$) with the typical wavelength of the GW emission. The calculations~are performed in the two regions --in principle-- independently. A {\it matching} procedure is required to read off the relevant parameters entering in the GW amplitude and phase in the radiation region (multipole moments) in terms of quantities from the near zone.\vskip 4pt

While the full computation within general relativity (GR) --using extended bodies and without separating into zones-- is devoid of UV divergences, there may be IR singularities, which are due to the long range nature of the Newtonian potential. For the computation of the GW amplitude, the IR sensitivity has two sources: The unknown value of the phase as the GW enters the detectors' band, and the resulting phase shift associated with scattering off of a long range potential. The former is absorbed into an initial conditions and the latter  cancels out in the total radiated power.\footnote{Let us emphasize that a cancellation occurs in principle for physical observables, 
and this may not be the case for gauge-dependent, unphysical, quantities.} On the other hand, the approximation scheme in which we split the calculation into regions generates new singular integrals. These additional divergences may be either of IR or UV origin. The latter appear as a consequence of  the point-particle approximation, whereas the former may include spurious poles which are not present in the full theory~calculations.\vskip 4pt 

The effective field theory (EFT) framework, coined Non-Relativistic General Relativity (NRGR) due to similarities with effective theories for the strong interaction \cite{nrgr}, is particularly useful to, not only organize the computation systematically, but also to elucidate the origin of IR and UV divergences and consistently eliminate them. For reviews of the EFT approach see \cite{walter,rafric,iragrg,rafgrg,riccardo,review}. In the EFT formalism the near/far zone separation is accomplished at the level of the action, by splitting the gravitational field into potential and radiation modes, such that each term scales homogeneously in $v$. Since all observables are computed in terms of Feynman diagrams, whose vertices and propagators are fixed by the effective action, it is a simple matter to determine which ones contribute at any given order in the expansion parameter.\vskip 4pt 

Near zone calculations, which  involve the potential field, may contain UV divergences due to the point-particle approximation, and can be removed by {\it counter-terms} that include all possible terms in the action compatible with the symmetries of the problem. In this sense, the removal of classical divergences is identical to renormalization in quantum field theories  \cite{markwalt,IraTasi}. It~can be shown that, up to order $(R/r)^5 \sim v^{10}$ (or 5PN), the only unknown parameters (in addition to $G_N$) are the masses of constituents, for non-rotating compact bodies. This is known as the effacement theorem \cite{damour}. In the EFT formalism, this result follows from the fact that the first term --beyond minimal coupling-- that one can write down in the worldline effective action is proportional to the square of the electric and magnetic components of the Weyl tensor \cite{nrgr}. The coefficients are known as the Love numbers \cite{love,iragrg}. (For spinning objects \cite{nrgrs}, on the other hand, finite size effects appear already at 2PN order \cite{prl,nrgrs1s2,nrgrs1s1}. We will only deal with non-rotating bodies in what follows.)  Thus, from the point of view of renormalization, there can be no physically relevant divergences until 5PN order, or in other words, no `ambiguity parameters' are needed until this~order. At~5PN, after the divergence is subtracted away, the remaining finite part is chosen such that it matches the physical value of the Love number for the underlying constituent. Moreover, the divergence must be UV in origin, since it arises from the short-distance region of the Feynman integral, which in the full theory is cut off by the size of the constituents. While the subtraction scheme is arbitrary, the physical results are not, and the independence from the choice of renormalization scale translates into the renormalization group evolution of the coefficients of the effective theory \cite{nrgr,review}. \vskip 4pt

In the far zone we also encounter UV divergences as a conseqence  of the multipole expansion in which the binary (as a whole) is also replaced by a point-like source. In that case, a counter-term is needed which is proportional to the (time-dependent) multipole moments of the binary \cite{andirad}. However, the  previously alluded to matching condition  takes care of the ambiguity in the remaining finite part of the (renormalized) multipoles, since these are obtained in terms of masses, positions, velocities, etc., of the constituents. (Because of the non-linearities of GR, the binding energy of the system also contributes to the source multipoles.) On other hand, because the effective theory is constructed to agree with the full theory in the IR, IR divergences in the far zone will be identical to the ones in the complete GR computation, and consequently cancel out  \cite{andirad,amps}.\vskip 4pt 

The near zone, on the other hand, may introduce spurious IR divergences if the split into regions is not handled with care. In principle, IR singularities cannot develop with potential modes. This would signal a region of momenta in the Feynman integrals which is much softer than what is allowed  --by definition-- in the near zone. In other words, the near region should be cut off where the far zone begins. Hence, the IR divergences in the computation of the binding potential cannot have a physical origin. The solution to this problem is to remove from the integrals in the near zone the contribution from Fourier modes with wavelengths larger than $r$. The procedure goes by the name of the {\it zero-bin} subtraction \cite{zero}. The purpose of this note is to demonstrate how to implement the latter in the PN framework, without the need of ambiguity parameters. 
The logic we are about to describe holds to all orders in the PN expansion.\vskip 4pt

We will implement the zero-bin subtraction using dimensional regularization (dim. reg.), for which the procedure becomes more transparent, due to the preservation of the symmetries and lack of additional momentum scales, unlike say a cut-off regulator. However, in principle the zero-bin subtraction is independent of the regularization scheme, and in particular it can be utilized also with the methods described in e.g. \cite{blanchet}.  As~an explicit example we will consider the conservative binding energy at 4PN order, where the spurious IR divergences in the near zone first appear \cite{4pnjs,4pnjs2,4pndjs,4pndjs2,4pnbla,4pnbla2,4pndj}. An illustrative case is presented in full in a companion paper, where similar considerations apply \cite{lamb}. 

\section{The zero-bin subtraction}

The calculation of the binding energy in full-fledged GR is finite, therefore the IR/UV divergences appear as a result of the approximation scheme. The above mentioned IR singularities that arises in near zone calculations is a signal of double-counting. Modes which are thought to be confined to the bound state are overlapping with the radiation region. Therefore, a consistent methodology must ensure that  spurious IR divergences are absent. In the EFT formalism such a methodology entails including ``zero-bin'' subtractions \cite{zero}.
We sketch the basic steps here, see \cite{zero} for more details.\vskip 4pt

The binding energy in NRGR is determined by convolutions of Green's functions 
of potential and radiation modes. Hence, if we denote by $V_{\rm full}$ the full gravitational potential, we have
\beq
V_{\rm full} = V_{\rm pot} + V_{\rm rad} \,,
\eeq
where $V_{\rm pot}$ is the potential mode contribution in NRGR, and $V_{\rm rad}$ is the conservative part of the tail corrections in the radiation theory. In principle, at a given $n$PN order we will have a result of the form
\beq
V^{(n)}_{\rm pot}(\bq) = \frac{A^{(n)}_{\rm pot}}{\epsilon_{\rm UV}} + \frac{B^{(n)}_{\rm pot}}{\epsilon_{\rm IR}} + f^{(n)}_{\rm pot}(\bq, c^{(n)}_i,\mu)\,,
\eeq
with $\bq \sim 1/r$, and  $\epsilon_{\rm UV/IR} \equiv (d-4)_{\rm UV/IR}$ in dim. reg., which we use to regularize the divergent integrals. The factors of $\mu$ appear also from dim. reg. since, in $d\neq 4$, Newton's constant acquires an extra mass scale proportional to $\mu^{d-4}$. The $A^{(n)}_{\rm pot}$ term can be either set to zero by a coordinate transformation for $n\leq 4$; or removed by a counter-term, $(c^{(n)}_{i})_{\rm c.t.} \propto 1/\epsilon_{\rm UV}$, using the coefficients of the effective action (beyond minimal coupling) for $n\geq 5$.\footnote{The divergences with $n\leq 4$ can be equally absorbed into counter-terms, which can be removed by field redefinitions \cite{nrgr}.} The finite part, encoded in $c^{(n)}_i(\mu)$, accounts for the finite size effects starting at 5PN order. The $\mu$-independence of physical results leads to a renormalization group evolution for these coefficients \cite{nrgr,review}. By explicit calculation it is known that $B^{(n)}=0$ for $n \leq 3$. On~the~other hand, the far zone contribution to the conservative part of the potential will be of the form\footnote{In principle we also need to incorporate the running of the multipole moments, e.g. $I^{ij}(\mu)$, see \cite{andirad}. However, these effects --which do not introduce ambiguities-- will enter at much higher orders. We suppress their contribution here for simplicity.}
\beq
V^{(n)}_{\rm rad}(\omega) = \frac{A^{(n)}_{\rm rad}}{\epsilon_{\rm UV}} + \frac{B^{(n)}_{\rm rad}}{\epsilon_{\rm IR}} + f^{(n)}_{\rm rad}(\omega,\mu)\,,
\eeq
with $\omega \sim v/r$. For our purposes we will take $B^{(n)}_{\rm rad}=0$, since IR divergences do not appear in the conservative contributions from the radiation sector. From the computation of the tail effect we find $A^{(n)}_{\rm rad} \neq 0$ for $n \geq 4$, and $A^{(n)}_{\rm rad} = - B^{(n)}_{\rm pot}$. This relationship is key in the cancellation of the spurious IR/UV divergences in the far/near zones from the conservative sector \cite{nltail}. The role of the zero-bin subtraction is to remove the IR divergent contribution from the potential region, and transform it into the required UV pole to cancel the one arising from the far zone. In other words, we must shift \beq V_{\rm pot} \to V_{\rm pot} - V_{\rm zero\text{-}bin},\eeq where $V_{\rm zero\text{-}bin}$ corresponds to an asymptotic expansion of the integral around the region responsible for the IR singularity. This procedure removes the double counting induced by the overlap between the IR-sensitive part of $V_{\rm pot}$ and the contribution from $V_{\rm rad}$.\vskip 4pt 

In dim. reg., the zero-bin part may involve a scaleless integral.These are usually set to zero, however, the procedure is more subtle for integrals which are simultaneously IR and UV divergent. That is because we need to isolate the UV and IR divergences, which otherwise could be incorrectly removed. This is a key aspect for the correct implementation of the zero-bin subtraction.
 For example, consider the following scaleless integral in the limit $d\to 4$, 
\beq
I=\int \frac {d^d k}{k^4}\,,
\eeq
which is clearly both IR and UV sensitive. One can easily manipulate the integrand and re-write it as IR and UV divergent parts, with $M$ some generic scale,
\beq
I = \int \frac {d^dk}{k^2-M^2} \frac{1}{k^2}-  \int \frac {d^dk}{k^2-M^2} \frac{M^2}{k^4}.
\eeq
To regularize the integral, in the first term we use a UV regularization, $\epsilon_{\rm UV}  < 0$, whereas in the second we chose $\epsilon_{\rm IR}>0$, leading to
\beq
I \propto  \left(\frac{1}{\epsilon_{\rm UV}}-\frac{1}{\epsilon_{\rm IR}}\right).
\eeq
The zero-bin is often of this form, i.e.
\beq
V^{(n)}_{\rm zero\text{-}bin} = B^{(n)}_{\rm pot} \left( \frac{1}{\epsilon_{\rm IR}} - \frac{1}{\epsilon_{\rm UV}}\right)\,,
\eeq
such that the IR part is precisely what we need to cancel the pole from the potential region.\footnote{Let us remark this is not a totally foreign procedure within dim. reg. See for example appendix B in \cite{andirad}, where scaleless integrals are essential to remove IR divergences in the computation of the tail contribution to the radiative multipole moments, and also to obtain the correct renormalization group equations.} At the same time, the remaining UV pole readily removes the divergence from the far zone. Hence, after the zero-bin is incorporated, we have
\beq
V^{(n)}_{\rm pot}-V^{(n)}_{\rm zero - bin} + V^{(n)}_{\rm rad} \to V^{(n)}_{\rm tot}(\bq,\omega, c^{(n)}_i(\mu),\mu)\,,
\eeq
where the $c^{(n)}_i(\mu)$ are the only parameters which require a matching computation beyond the PN framework, first entering at 5PN order according to the effacement theorem \cite{damour,nrgr,review}. For the case $n\leq 4$, the final expression is not only finite, but also all the $\log \mu$ terms --associated to either IR or UV poles-- cancel out (or are removed by a coordinate transformation). This is the situation we encounter at 4PN order, we describe below. Before we move on, let us emphasize an important point. While using dim. reg. (for both IR and UV divergences) the zero-bin subtraction may turn out to be algebraically trivial, as in the example at hand, that will not be the case with other regularization procedures or more complicated scenarios. Moreover, even in dim. reg. the zero-bin could involve non-trivial momentum dependence, see for \cite{forwardRS} for an example of this sort.

\section{The conservative dynamics at 4PN order}

The complete conservative dynamics of a gravitationally bound (non-spinning) two-body system has been recently determined at 4PN order in \cite{4pnjs,4pnjs2,4pndjs,4pndjs2,4pnbla,4pnbla2,4pndj}. However, the calculations thus far have been plagued by ambiguity parameters. Unlike previous ambiguities at 3PN order, which are  due to UV singularities and can be removed by coordinate transformations \cite{blanchet} (or field redefinitions in the EFT language \cite{nrgr,review}), on this occasion the ambiguities are due to the scheme-dependence introduced in \cite{4pnjs,4pnjs2,4pndjs,4pndjs2,4pnbla,4pnbla2,4pndj} to handle the presence of IR divergences. Hence, the exact form of the analytic contribution to the gravitational binding energy at leading order in the symmetric mass-ratio, $\nu= \tfrac{m_1m_2}{(m_1+m_2)^2}$, has been derived after a comparison with gravitational self-force results was made \cite{4pndjs,4pnbla2}, see also \cite{ALT2}. However, one should not have to rely on calculations outside the scope of the PN expansion to remove spurious divergences which are due to the approximation scheme. (If for no other reason that it requires unnecessary extra work.) In fact, this issue will re-emerge at higher PN orders, involving IR-sensitive contributions at ${\cal O}(\nu^k)$, with $k\geq 2$, for which self-force computations may not be available. The zero-bin subtraction \cite{zero}, on the other hand, removes the need of ambiguity parameters altogether, yielding IR-safe quantities. We illustrate the procedure for the calculation of the 4PN gravitational potential below. 

\subsection{The ambiguities}

In \cite{4pnjs,4pnjs2}, where the calculation was performed using dim.~reg. in the Arnowitt, Deser, and Misner (ADM) formalism, the regularized gravitational potential contains an IR pole,\footnote{As it turns out, the derivation in \cite{4pnjs,4pnjs2} uses a combination of regularization methods before matching the calculation into the far zone later on in \cite{4pndjs}. This exacerbates the ambiguous nature of the computation.} 
\beq
\label{eq:V4pn1}
\int dt \, V^{ JS}_{\rm 4PN} (t,\mu) = -\frac{G_N^2 M}{5} \int \frac{d\omega}{2\pi} \omega^6 I^{ij}(-\omega) I^{ij}(\omega) \left(\frac{1}{\epsilon_{\rm IR}} - 2\log (\mu r)\right) + {\rm local/finite} \,.
\eeq
We may attempt to remove the IR divergence from the near zone calculation, by subtracting it away. However, as it was originally understood by the authors of \cite{4pnjs,4pnjs2},  this is totally arbitrary since one could subtract either $1/\epsilon_{\rm IR}$ or $1/\epsilon_{\rm IR}+$constant. In general, different subtraction schemes, or even choices of regulators,  
e.g. graviton-mass, momentum cut-off, Hadamard, etc., would produce a non-unique result. Therefore, the final answer would be ambiguous and the 4PN near zone Hamiltonian would depend on a regularization scale, $\mu$, as well as an undetermined dimensionless constant, $C$, through (the factor of $2/5$ is a convention)
\beq
V^{JS}_{(C)} \equiv  \frac{2}{5} C\, G^2_N M \left(I^{ij(3)}(t)\right)^2\,.
\label{H4pnc}
\eeq 
The same type of ambiguities (in principle more than one) were introduced in the computation of the Fokker-action in harmonic gauge \cite{4pnbla,4pnbla2}. The existence of these extra parameter(s) was also discussed in more detail in\cite{4pndjs2}. In the ADM and Fokker-action approaches, the value of $C$ is fixed after importing knowledge from an independent calculation. Since the ambiguity shows up at leading order in $\nu$, the constant $C$ (in principle more than one) was fixed by comparison with existent self-force calculations \cite{4pndjs,4pnbla2}.\vskip 4pt  

On the other end, we have the contribution to the gravitational potential at 4PN order from radiation-reaction, which was computed within the EFT approach in~\cite{nltail} (see also \cite{tailfoffa}), yielding
\beq
\int dt\, V_{\rm tail}(\mu) =    \, \frac{G_N^2 M}{5} \int \frac{d\omega}{2 \pi} \,  \omega^6 \, I^{ij}(-\omega) I^{ij}(\omega) \left[  \frac{1}{\epsilon_{\rm UV}} + \log \frac{\omega^2}{\mu^2} + \gamma_E - \log\pi - \frac{41}{30} \right] \,.
\label{eq:RRnl}
\eeq
We notice that, instead of IR issues, we encounter a UV divergence. The alert reader will immediate realize that the IR and UV near/far zone poles in \eqref{eq:V4pn1} and \eqref{eq:RRnl} have --up to a crucial sign-- the same coefficients \cite{nltail}, provided the IR divergence from \cite{4pnjs,4pnjs2} is reproduced in NRGR (see below).\footnote{As emphasized in \cite{nltail}, this must be the case since the factors for the pole and logarithm are related in dim. reg., and the coefficient for the latter is fixed by the long-distance contribution to the binding energy, first obtained in \cite{ALTlogx} and re-derived in \cite{nltail} (see also \cite{andirad3}).} In our language, an expression like \eqref{H4pnc} would arise if we were to use two different subtraction scales for the IR and UV poles (i.e. $\mu_{\rm IR}$ and $\mu_{\rm UV}$), leading to an ambiguity parameter, $C \propto \log (\mu_{\rm UV}/\mu_{\rm IR})$ \cite{nltail}; however, this would be inconsistent.\footnote{The dim. reg. prescription must be universally applied in both regions (regardless of whether we have IR or UV divergent integrals) with a $d$-dimensional $G_N$ entering in the Einstein-Hilbert action.} We~show in what follows how to unify the treatment of the potential and radiation singularities in the EFT formalism. 

\subsection{Infrared divergences in the EFT framework}

\begin{figure}[t!]
\centerline{{\includegraphics[width=0.5\textwidth]{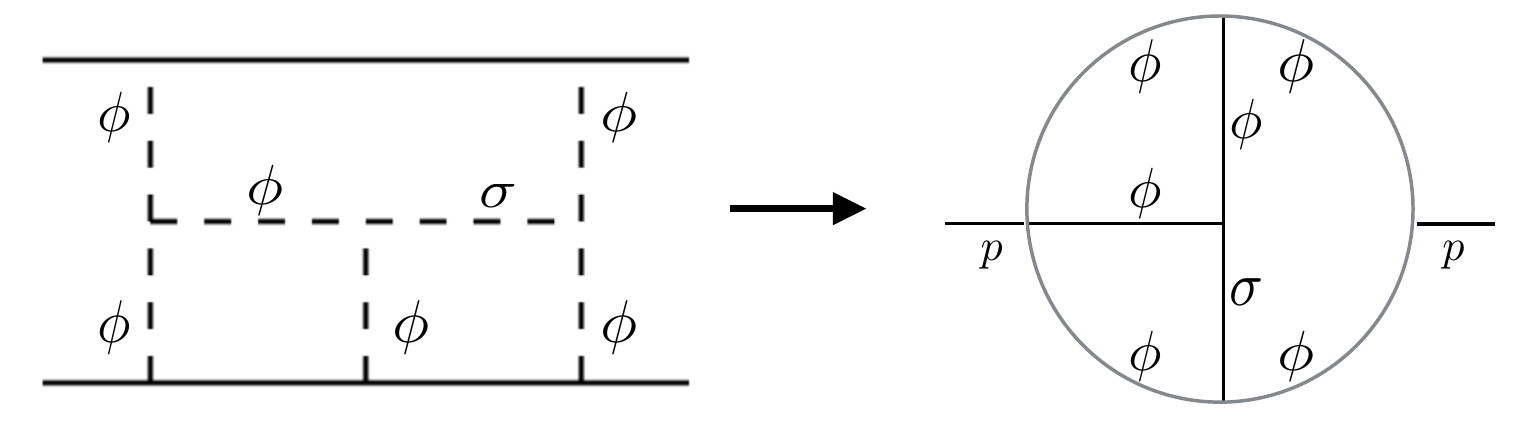}}}
\caption[1]{A contribution in NRGR to the gravitational potential from the near zone at 4PN order. We use the decomposition introduced in \cite{KK} in terms of $(\phi,\bA_i,\gamma_{ij})$ fields.  As described in \cite{nrgr4pn2}, the diagram may be cast as a `three-loop' self-energy amplitude in gauge theory, which can be further decomposed in a series of master integrals.} \label{3loopot}
\end{figure}

While the UV pole in the conservative sector from the tail effect  identified in \cite{nltail} was expected, the IR divergences in the potential region have not been yet fully isolated in NRGR. Hence, we must first identify the terms from the near zone which yield IR poles at 4PN order. There are many diagrams which contribute to the gravitational potential at this order \cite{nrgr4pn1,nrgr4pn2}. One such term is depicted in fig.~\ref{3loopot}. Following the representation in \cite{nrgr4pn2}, the integral may be written as
\beq
{\rm fig.}\,\ref{3loopot} = m_1^2m_2^3 \int_{\bk_1,\bk_2,\bk_3} \frac{{\rm Num}(\bk_i,\bp)}{\bk_2^2\,\bk_3^2\, \bk_{12}^2\,\bk_{13}^2\,\bk_{23}^2\,\bp_2^2\,\bp_3^2}\label{3loo}\,,
\eeq
where $ \int_{\bk_i} \equiv \int \frac{d^{d-1}\bk_i}{(2\pi)^{d-1}}$, $\bp_i \equiv \bp-\bk_i$ and $\bk_{ij} \equiv \bk_i - \bk_j$. The numerator, ${\rm Num}(\bk_i,\bp)$, depends on the loop momenta, $\bk_i$, as well as the Fourier variable, $\bp$, related to the relative distance, $\br \equiv \bx_1-\bx_2$. In principle this topology contributes at ${\cal O}(G_N^4)$, and naively it would have entered also at 3PN order. However, the $\phi^3$ vertex introduces an extra factor of $v^2$ \cite{KK}. Therefore, the diagram in fig.~\ref{3loopot} enters at order $G_N^4v^2$, namely~4PN. The calculation can be reduced to a series of master integrals, in particular one such integral will be of the form
\beq
\label{masterM}
{\cal M}(\bp) = \int_{\bk_1,\bk_2,\bk_3} \frac{1}{\bk_1^2 \bk_{2}^2\bk_3^2 \bd^2_2 \bd^2_3}\,,
\eeq
with $\bd_2 = \bp+\bk_1-\bk_2,\, \bd_3=\bp+\bk_1-\bk_2-\bk_3$, so that the contribution to the gravitational potential is given by,
\beq
\label{FourM}
V_{{\rm fig.}\ref{3loopot}} = c_{\cal M} m_1^2m_2^3 \int_\bp \bp^2\, {\cal M}(\bp) e^{i\bp\cdot \br}+ \cdots\,,
\eeq
with some numerical pre-factor, $c_{\cal M} \propto G_N^4$. \vskip 4pt

The master integral may be computed in $d$-dimensions and is IR divergent for $d\to 4$. The relevant IR divergent part is given 
by\footnote{See \cite{Gehrmann} for the complete expression which includes various $\Gamma$-functions.}
\beq
{\cal M} \propto |\bp|^{-1+\frac{3}{2}\epsilon_{\rm IR}} \, \Gamma[\epsilon_{\rm IR}] \cdots .
\eeq
The origin of this divergence can be traced down to the standard formula,
 \begin{align}
 \label{footl}
 I[n_1,n_2]&\equiv \int_\bk \frac{1}{\left[[\bk^2]^{n_1}(\bk+\bp)^2\right]^{n_2}} \\ &= \frac{\Gamma[n_1+n_2-(d-1)/2]\Gamma[(d-1)/2-n_2]\Gamma[(d-1)/2-n_1]}{\Gamma[n_1]\Gamma[n_2]\Gamma[(d-1)-n_1-n_2]} \left(\bp^2\right)^{(d-1)/2-n_1-n_2}\,,\nn
 \end{align}
which can be used repeatedly (three times) to factorize the integral. The IR divergence is manifest by the fact that at some point a term may become singular for $d<4$. 

\subsection{Removing the double-counting}

The implementation of the zero-bin subtraction for the master integral in \eqref{masterM} is relatively straightforward in dim. reg. Because it can be factorized, the IR pole can be isolated from (see \eqref{footl})
\beq
I\left[\frac{3-\epsilon}{2},\frac{1-\epsilon}{2}\right] \propto |\bp|^{-1 + \frac{3}{2} \epsilon} \Gamma[\epsilon],
\eeq
which contains the expected term $\propto 1/\epsilon_{\rm IR}$  coming from the region $\bk \ll \bp$, which must be subtracted away. Hence, expanding the factor of $(\bk+\bp)$, we find
\beq
I_{\rm zero\text{-}bin} \left[n_1,n_2\right] = \int_\bk \frac{1}{[\bk^2]^{n_1}[\bp^2]^{n_2}} \xrightarrow{(n_1\to 3/2,n_2 \to 1/2)} 
|\bp|^{-1} \int_\bk \frac{1}{\bk^3} \propto |\bp|^{-1} \left(\frac{1}{\epsilon_{\rm UV}} - \frac{1}{\epsilon_{\rm IR}}\right)\,. \label{zerobin0}
\eeq
Notice the zero-bin subtraction removes the IR pole, essentially transforming it into a UV singularity. This example illustrates how the procedure works at the level of the master integrals.\footnote{Let us emphasize once again that one cannot set to zero these type of integrals when IR divergences are present \cite{IraTasi}. Moreover, they are key to obtaining the correct renormalization group evolution for the Wilson coefficient in the effective action, e.g. \cite{andirad}. Hence, the zero-bin subtraction not only properly removes the IR divergences in the conservative sector, it will also play a crucial role at 5PN and higher orders to compute the renormalization group equations for the finite size terms in the effective theory.}\vskip 4pt

Plugging the expression for the master integral into \eqref{FourM}, including the zero-bin subtraction, we get a term,
\beq
\label{vfig1}
V_{{\rm fig.}\,\ref{3loopot}} \to \frac{ G^4_N m_1^2m_2^3 v^2}{r^4} \left(-\frac{1}{\epsilon_{\rm UV}} + 2 \log \mu r\right) +\cdots\,,
\eeq
where the IR pole turned into a UV divergence. Using the equations of motion (EOM), we have
\beq
G_N^2 (m_1+m_2) (I^{ij(3)})^2\, \xrightarrow{\rm EOM}\, \frac{G_N^4 \left(m_1^2m_2^3 + m_2^2m_1^3\right)v^2}{r^4} \big(1 + {\cal O}(v^2)\big) \,,
\eeq
such that the contribution from fig.\,\ref{3loopot} to the gravitational binding potential has the correct structure to cancel the UV divergence from the conservative part of the tail effect (and $\log\mu$ factor), as advertised \cite{nltail}. The zero-bin subtraction must be implemented in all of the IR-sensitive master integrals that enter at ${\cal O}(G_N^4v^2)$. Notice that, while we referred to the specific contribution from fig.~\ref{3loopot}, the argument applies more generally to  the integral in \eqref{masterM}, and ultimately the one in \eqref{footl}. The latter will enter in the decomposition of many of the three-loop amplitudes, and therefore IR poles may appear in principle from a series of Feynman diagrams, not just the one in fig.~\ref{3loopot}. In fact, it turns out all of the IR-sensitive contributions, from the master integrals entering in the full computation at 4PN, can be reduced to the one-loop integral in \eqref{footl}.\footnote{This has been shown to be the case in the (soon to appear) complete computation using the EFT approach at 4PN order, following the methodology discussed in \cite{nrgr4pn2} and integration-by-parts. (Notice there are no IR divergences at ${\cal O}(G_N^5)$ \cite{nrgr4pn2}.) We thank Riccardo Sturani (private communication) for confirming this to us.} After the zero-bin subtraction, the result will take the form in \eqref{vfig1} simply from dimensional analysis.\vskip 4pt

The reader may wonder about the above procedure if momentum cut-offs were invoked to regularize the divergences in \eqref{footl}, for example adding a small mass in the IR and using Pauli-Villars regularization in the UV. In such scenario --which is not the one advocated here since it breaks the symmetries of the theory-- the contribution from the zero-bin integral would read:
\beq
\label{eqpv}
 \int_\bk \frac{1}{[\bk^2-m_g^2]^{3/2}[\bp^2]^{1/2}}- \int_\bk \frac{1}{[\bk^2-\Lambda^2]^{3/2}[\bp^2]^{1/2}}  \propto |\bp|^{-1} \Big( \log m_g - \log \Lambda\Big)\,,
\eeq
instead of \eqref{zerobin0}. The $\log m_g$ term would cancel out against a similar IR singularity in the original integral prior to the zero-bin subtraction, whereas the $\log\Lambda$ removes the UV pole from the tail contribution \cite{nltail}, which must be consequently also regularized using a Pauli-Villars cut-off. Notice there is no sense in which \eqref{eqpv} can be set to zero in this framework. We should stress, however, that in general breaking diffeomorphism invariance (also adding a graviton mass) introduces several other problems which we refrain from discussing in the present work.

\section{Concluding remarks}

The gravitational binding energy contains a logarithmic contribution, $E_{\rm log} \propto v^8 \log v$, in GR. When performing PN calculations, using all the present methodologies, one splits the problem up into near and far zone, which leads to both IR and UV logarithmically divergent integrals. The sum of these results reproduces
the aforementioned logarithm, but at the cost of introducing a set of spurious new poles. While the UV divergences are well understood, the presence of IR divergences is a new development in the PN framework, which first entered scene at 4PN order for non-rotating bodies.\footnote{Up until now, the computations within the EFT approach had not encountered IR divergences in the potential region, and the only IR poles in the radiation theory arose from the tail contribution to the one-point function (radiative multipole moments), previously  mentioned. There are other  UV divergences in the radiation sector, which are responsible for the renormalization group evolution for the multipole moments (or can be consistently set to zero in dim. reg.), e.g. \cite{review}.} The IR sensitivity in the near region has led to the introduction of ambiguity parameters in the methodology implemented in \cite{4pnjs,4pnjs2,4pndjs,4pndjs2,4pnbla,4pnbla2,4pndj}. This occurs in any framework in which the IR/UV divergences, from the near/far zone contributions to the gravitational potential, are treated independently. That is what happens in the computations in \cite{4pnjs,4pnjs2,4pndjs,4pndjs2,4pnbla,4pnbla2}, which removed the IR singularity in the near zone prior to appending the contribution from the far region, obtained independently in~\cite{tail3,tail3n}. However, by implementing the zero-bin subtraction there are no ambiguity parameters at any stage of the calculation. The procedure removes the near zone IR divergences, transforming them into UV poles which readily cancel the singularities from the far region. We illustrated the zero-bin subtraction in the calculation of the conservative dynamics at 4PN order using dim. reg., in which case it  replaces $\epsilon_{\rm IR} \to \epsilon_{\rm UV}$ by subtracting from the near region calculation a series of scaleless integrals. This procedure justifies the claims in \cite{nltail} that the sum of the near/far region IR/UV poles in dim. reg. must cancel each other out. However, the formalism is more general and may be also implemented for other regularization schemes, where a cancellation of IR and UV divergences may be far less obvious, including the ones in \cite{4pnjs,4pnjs2,4pndjs,4pndjs2,4pnbla,4pnbla2,4pndj}. (See \cite{zero} for a discussion of the zero-bin subtraction in the confines of a cut-off regulator.) The virtue of our procedure is precisely the independence from the regulator and subsequent lack of ambiguities to all orders in the PN expansion. The zero-bin subtraction not only provides finite results, it also uniquely fixes the analytic (local) contribution to the potential without the need of extra matching conditions beyond the PN framework.  An explicit example is described in \cite{lamb} in the context of electrodynamics.\vskip 4pt
 
{\bf Note added:} In a recent paper \cite{BlanchetDR} it was shown how the implementation of dim. reg. for both IR/UV divergences (as advocated here) removes an ambiguity parameter within the formalism of \cite{4pnbla,4pnbla2}, also confirming our original claim that the near and far zone divergences (as well as factors of $\mu$) cancel each other out without the need of extra information \cite{nltail}. As it was argued in \cite{nltail}, the cancelation in \cite{BlanchetDR} is achieved by identifying $\epsilon_{\rm IR} \leftrightarrow \epsilon_{\rm UV}$ which, as we discussed here, is formally justified by the zero-bin subtraction. Moreover, it was noted that the remaining ambiguity parameter still present in \cite{BlanchetDR} is automatically fixed within the EFT framework --with the correct coefficient previously obtained in \cite{nltail}, i.e. $2\kappa =41/30$ (see \eqref{eq:RRnl} above and Eq.~5.6 in \cite{BlanchetDR})-- provided the completion of the local part of the 4PN potential, initiated in \cite{nrgr4pn1,nrgr4pn2}, agrees with the derivation in \cite{4pnbla,4pnbla2}, as it is expected in harmonic gauge. 
 
 \section*{Acknowledgments}
We~thank the participants of the workshop `Analytic Methods in General Relativity' held at ICTP-SAIFR,\footnote{\small \url{http://www.ictp-saifr.org/gr2016}} (supported by the S\~ao Paulo Research Foundation (FAPESP) grant 2016/01343-7) for fruitful interactions, in particular to Luc Blanchet, Guillaume Faye and Gerhard~Sch\"afer. We thank also Aneesh Manohar for discussions. R.A.P. thanks the theory group at DESY (Hamburg) for hospitality while this paper was being completed. The work of R.A.P. is supported by the Simons Foundation and FAPESP Young Investigator Awards, grants 2014/25212-3 and 2014/10748-5. I.Z.R. is supported by the National Science Foundation under grant NSF-1407744. 
 
\bibliographystyle{utphys}
\bibliography{RefNote}

\end{document}